\documentclass[12pt]{article}
\usepackage{epsf}
\def\dst{\displaystyle}
\def\f{\frac}

\def\m{\mu}

\def\t{\theta}
\def\p{\partial}
\def\e{\varepsilon}
\def\vp{\varphi}
\def\z{\zeta}
\def\eps{\varepsilon}
\def\be{\begin{equation}}
\def\ee{\end{equation}}
\def\bea{\begin{eqnarray}}
\def\eea{\end{eqnarray}}
\def\a{\alpha}
\def\ba{\begin{array}}
\def\ea{\end{array}}
\def\bea{\begin{eqnarray}}
\def\eea{\end{eqnarray}}

\def\l{\left}
\def\r{\right}

\begin{document}
\begin{titlepage}
\begin{center}
{\Large \bf William I. Fine Theoretical Physics Institute \\
University of Minnesota \\}
\end{center}
\vspace{0.2in}
\begin{flushright}
FTPI-MINN-08/40 \\
UMN-TH-2722/08 \\
October 2008 \\
\end{flushright}
\vspace{0.3in}
\begin{center}
{\Large \bf Spontaneous decay of a metastable domain wall \\}
\vspace{0.2in}
{\bf A. Monin \\}
School of Physics and Astronomy, University of Minnesota, \\ Minneapolis, MN
55455, USA, \\
and \\
{\bf M.B. Voloshin  \\ }
William I. Fine Theoretical Physics Institute, University of
Minnesota,\\ Minneapolis, MN 55455, USA \\
and \\
Institute of Theoretical and Experimental Physics, Moscow, 117218, Russia
\\[0.2in]
\end{center}

\begin{abstract}

We consider the decay of a metastable domain wall. The transition proceeds
through quantum tunneling, and we calculate in arbitrary number of dimensions
the preexponential factor multiplying the leading semiclassical exponential
expression for the rate of the process. We find that the effect of the motion in
transverse directions reduces to a renormalization of the tension of the edge of
the wall in the semiclassical exponent. This behavior is similar to the one
previously found for breaking of a metastable string. However this simple
property is generally lost for spontaneous decay of higher-dimensional branes.

\end{abstract}

\end{titlepage}

\section{Introduction}
\begin{figure}[ht]
  \begin{center}
    \leavevmode
    \epsfxsize=5cm
    \epsfbox{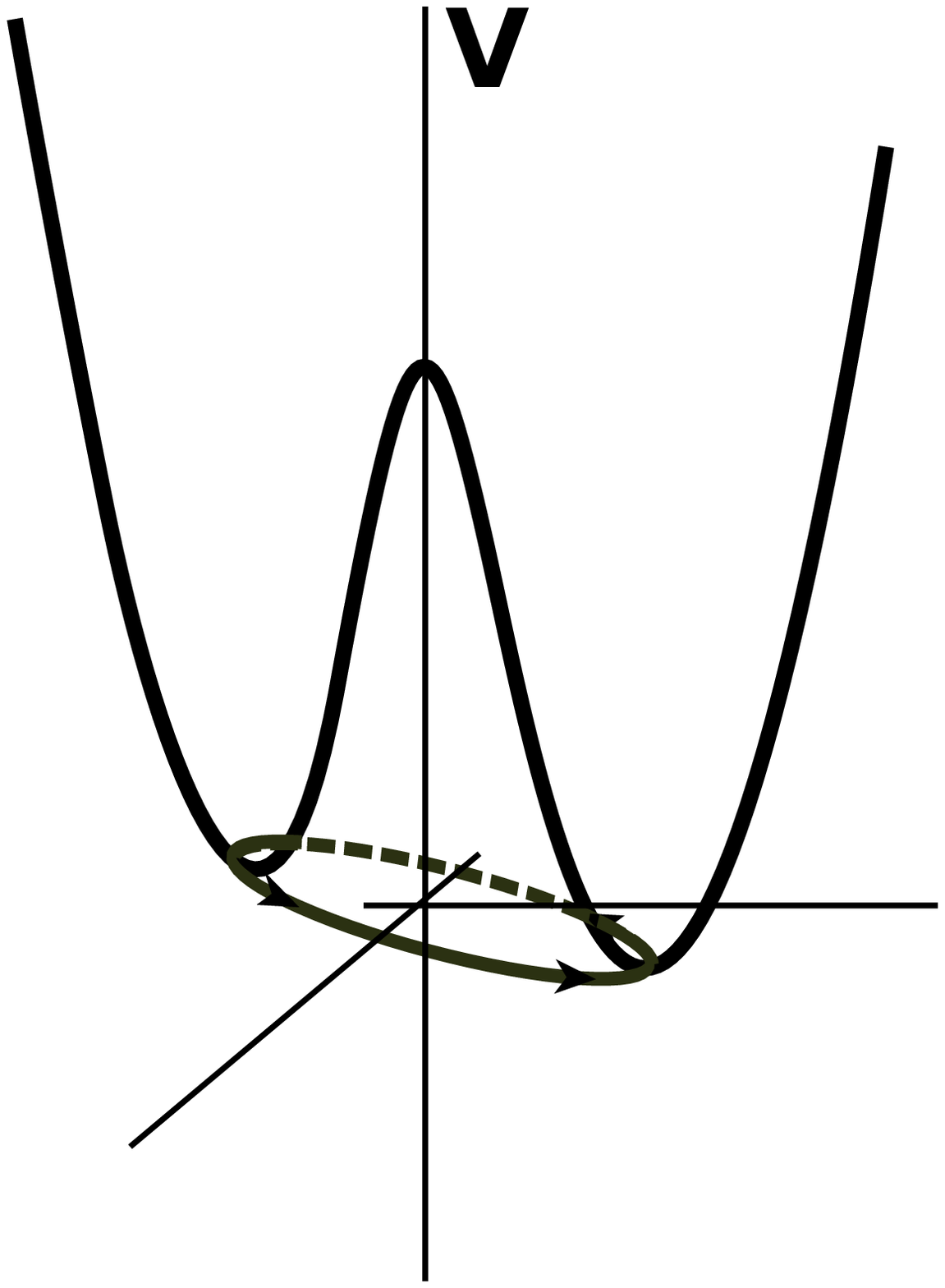}
    \caption{Potential}
  \end{center}
\label{potential}
\end{figure}

Metastable domain wall solutions arise in models with spontaneously broken
approximate symmetry. The existence of such solutions was shown from different
points of view, for example, in \cite{shifman97,witten98a,witten98b}. The origin
of a metastable wall can be illustrated in a model with potential shown in
Fig.\ref{potential}, it corresponds to the interpolation between the same vacuum
state at two spatial infinities, e.g. at $z=-\infty$ and $z=+\infty$ with the
field winding around the `peg' in the potential.

Such configuration is classically stable in the sense that the solution with
certain topological number (winding number) can not evolve classically to a
solution with different topological number. However, such a transition can
proceed due to temperature fluctuation, when the path, corresponding to the
solution can be lifted over the barrier or due to quantum tunneling (the motion
in a classically forbidden region), hence the domain wall can undergo a decay at
certain conditions. In this paper we restrict ourselves to the later case.

The decay of a domain wall is analogous to decay of metastable
vacuum\cite{vko,Coleman,mv2004} in $2+1$ dimensions. Indeed, if a hole with area
$A$ is created in a wall, the gain in the energy is $\eps \, A$. The barrier, on
the other side, that inhibits the process is created by the energy $\mu \ell$,
with $\ell$ being the perimeter of the hole and $\m$ being a tension associated
with the interface. Thus, the `area' energy gain exceeds the barrier energy only
starting from a critical size of the hole created, i.e. starting with a round
hole of radius $R=2 \mu/\eps$. Once a critical piece has nucleated due to
tunneling, it expands, converting the domain wall. Therefore the probability of
the transition is given by the rate of nucleation of the critical holes in the
domain wall. The important difference, addressed in this paper, from the $2+1$
dimensional false vacuum decay is that the domain wall can move in the
transverse direction(s) and this motion affects the nucleation rate at the level
of the preexponential factor.

Recently we have considered a similar problem of calculating the rate of a
transition between two states of a string with different tension~\cite{mv08}. It
was shown that in a case of complete breaking (transition into nothing) of a
string with tension $\e$ the essential effect of the motion in transverse
dimensions reduces to a renormalization of mass parameter $\m$ associated with
the interface between two states, such that the dependence of the rate on
string's tension has the same form as for two dimensions, where it coincides
with well-known one for false vacuum decay or Schwinger process of charged pair
production in external electromagnetic field \cite{Schwinger}, expressed in terms of renormalized
parameter $\m_R$
\be
\f { d \Gamma } { d \ell }  = \f {\e} {2 \pi} \exp \l( - \f {\pi \m_R^2} {\e}
\r).
\ee
As a result of the calculation to be presented in this paper we find that a
similar behavior also applies to the decay of a metastable wall. Namely the
essential effect of the motion of the wall in the transverse dimension(s)
reduces to a renormalization of the boundary tension parameter $\m$ in the
expression for the false vacuum decay in $2+1$ dimensions. It can be noted that
such a reduction is not trivial and holds only for the transitions of strings
and two-dimensional walls. We have explicitly verified that such behavior is
lost in similar transition of higher dimensional branes, where the transverse
motion produces effects that are not reduced to a renormalization of the tension
of the interface.

The decay rate of a QCD domain wall with tension $\e$ with further creation of
the interface with tension $\m$ was considered in Ref.~\cite{fzh} by adapting
the expression for metastable vacuum decay in $2+1$ dimensions:
\be
\f { d \Gamma } { d A }  = \mathcal{P} \exp \l( - \f {16 \, \pi \, \m^3} { 3 \,
\e^2 } \r),
\ee
where $\mathcal{P}$ is the preexponential factor. This factor is found from a
calculation of the path integral over small deviations from the semiclassical
tunneling trajectory. In the limit of small $\eps$ the result of such
calculation in a (2+1) dimensional theory, can be readily copied from the
corresponding expressions in the equivalent (2+1) dimensional problem of false
vacuum decay\cite{mv2004}
\be
\mathcal {P} _ { d = 2+1 } =\f { \mathcal{C} } { \e ^ {7/3} },
\label{prevac}
\ee
with $\mathcal{C}$ being a constant independent of wall tension $\e$, which can
not be found from effective description of the theory. In other words, the
constant $\mathcal{C}$ is determined by details of the underlying `microscopic'
model, but the dependence of the factor (\ref{prevac}) on $\eps$ is universal.

In this paper we calculate the domain wall decay rate per unit area in arbitrary
number of dimensions $d$, and find the result in a similar form,
\be
\f {d \Gamma} {d A} = \f {\tilde{\mathcal {C}}} {\e ^ {7/3}} \, \exp \l( - {16
\, \pi \, \m _ R ^ 3 \over 3 \, \eps^2} \r) ~,
\label{rate}
\ee
where again the constant $\tilde{\mathcal {C}}$ does not depend on $\eps$, and
$\m_R$ is the tension of the interface that includes the renormalization effect
of the wall fluctuations in the transverse dimensions.

The material in this paper is organized as follows. In Sec.~2 we formulate the
problem in terms of the effective Euclidean-space action, and in Sec.~3 we
consider the separation of variables in the relevant path integrals with this
action and also find formal expressions for these integrals in terms of
products. A Pauli-Villars regularization for the products is introduced in
Sec.~4, and the actual calculation is done in Sec.~5. In Sec.~6 we demonstrate
that the dependence of the result on the regulator parameter is completely
absorbed into the renormalization of the interface tension $\mu$. Finally in
Sec.~7 we present our results and conclusions.

\section{Euclidean action}
\begin{figure}[ht]
  \begin{center}
    \leavevmode
    \epsfxsize=7cm
    \epsfbox{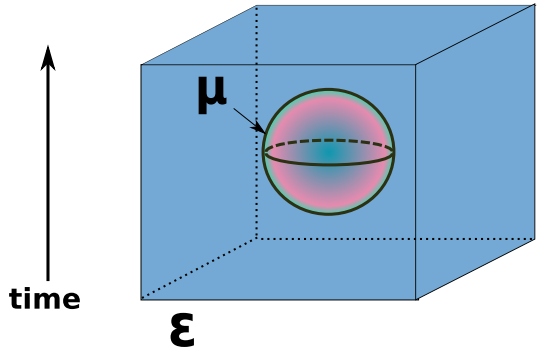}
    \caption{Bounce configuration}
  \end{center}
\label{wall_classical}
\end{figure}

The low energy effective action for the problem at hand is given by Nambu-Goto
action for two and tree dimensional objects
\be
S = \m \, \mathcal {A} + \e \, V,
\label{NG}
\ee
with $V$ being the world volume of the wall, while $\mathcal {A}$ is the world
area of interface. Nontrivial classical solution (bounce), which defines the
exponential behavior of the rate is empty sphere with radius
\be
R = \f {2 \m} {\e},
\ee
surrounded by the metastable phase (Fig.\ref{wall_classical}).
The action (\ref{NG}) is the low energy effective expression, and it does not
take into account the thickness of the wall, or of the interface. Hence it is
valid only while one can neglect the structure of the objects and consider
them as having zero thickness. If the thickness of the wall is of order $r_0$
then the
natural mass scale associated with it is $M_0=1/r_0$. Therefore one can write
the conditions of applicability of the action (\ref{NG})

\be
k \ll M_0, ~~~~ \ell \gg r_0,
\ee
with $k$ and $\ell$ being any momentum and length scales in the problem. For
instance the radius of the bubble should be much larger then the thickness of
the wall
\be
M_0 R = \f {2 \m M_0} {\e} \gg 1.
\ee
The probability rate of the transition is given by the imaginary part of the
ratio of the partition functions calculated around the bounce and the trivial
configurations
\be
{d \Gamma \over d A}= {1 \over AT} \, \mathrm{Im}\f{{\cal
Z}_{12}}{{\cal Z}_{1  }}~.
\label{rz}
\ee
The imaginary part of ${\cal Z}_{12}$ arises from one negative mode at the
bounce configuration. Furthermore, due to three translational zero modes, the
numerator in Eq.(\ref{rz}) is proportional to the total world volume $AT$
occupied by the wall,
so that the finite quantity is the transition probability per unit time (the
rate) and per unit area of the
wall.

In order to evaluate the relevant path integrals with the pre-exponential
accuracy we use the generalized cylindrical coordinates, with $r$, $\theta$ and
$\varphi$ being the
spherical variables in the $(t,x,y)$ space (of the bounce), and $z$ being the
transverse coordinate. We consider only one transverse coordinate, since the
effect of each of the extra dimensions factorizes, so that the corresponding
generalization is straightforward. We further assume, for definiteness, that the
space-time boundary in the $(t,x,y)$ space is a sphere of large radius $L$,
where
the boundary condition for the wall is $z(r=L)=0$. The small deviations of the
wall configuration from the bounce can be parametrized
by the variation of the bulk: $z(r,\theta,\varphi)$, the radial ($f$) and the
transverse ($\zeta$) shifts of the boundary:
\be
r(\theta,\varphi) = R +
f(\theta,\varphi)\,~~~~z(R,\theta,\varphi)=\zeta(\theta,\varphi)~,
\label{bv}
\ee
In terms of these variables the action (\ref{NG}) can be written in the
quadratic approximation in the deviations from the bounce as
\bea
&\dst S_{12}=\f {4 \pi} {3} \eps \, L^3  + {16 \, \pi \, \mu^3 \over 3 \,
\eps^2} ~~ + & {\m \over 2} \, \int d \Omega \, \left (g^{ab} \p_a f \p_b f-2 \,
f^2+g^{ab} \p_a \zeta \p_b \zeta
\right )+
\nonumber \\
&& {\e \over 2} \, \int_R^L r^2 \, dr \,d \Omega \, g^{ij} \, \p_i z \, \p_j z~,
\label{s12}
\eea
where the tensor $g_{ij}$ is the Euclidean metric tensor in spherical
coordinates ($i,j=r,\t,\vp$), while
$g_{ab}$ is induced metric on the sphere ($a,b=\t,\vp$), $d \Omega = \sin\t \,
d\t \,d\vp$.

The action around a trivial configuration in the quadratic
approximation takes the form
\bea
&& S_{1}=\f {4 \pi} {3} \eps \, L^3  + {\e \over 2} \, \int_0^L r^2 \, dr \,d
\Omega \, g^{ij} \, \p_i z \, \p_j z~,
\label{s1}
\eea

\section{Separating variables in path integral}
One can immediately see that the variable corresponding to the longitudinal
variations $f$ of the boundary is decoupled from other variables. This implies
that the path integral over $f$ can be considered independently of
the integration over other variables and that it enters as a factor in ${\cal
Z}_{12}$. On the other hand it is this integral that provides the imaginary part
to the partition function, and it is also proportional to the total space-time
area $AT$. Moreover, this path integral is identical to the one entering the
problem of false vacuum decay in (2+1) dimensions and we can directly apply the
result of that calculation \cite{mv2004}:
\be
{1 \over AT} \, \mathrm{Im} \int {\cal D} f \, \exp
\left [ {\m \over 2} \, \int d \Omega \, \left (g^{ab} \p_a f \p_b f-2 \, f^2
\r) \right]
=  {\mathcal{C} \over \e ^ {7/3} }~,
\label{2+1d}
\ee
where $\mathrm{C}$ is independent of $\e$ and it depends on parameters of
underlying theory.
The expression for the transition rate thus can be written in the form
\be
{d \Gamma \over d A}={\mathcal{C} \over \e ^ {7/3} } \, \exp \l( - {16 \, \pi \,
\mu^3 \over 3 \, \eps^2} \r) \,
{{\tilde {\cal Z}}_{12} \over {\cal
Z}_1} ~,
\ee
with the path integral ${\tilde {\cal Z}}_{12}$ running only over the transverse
variables $\zeta$ and $z$
\be
{\tilde {\cal Z}}_{12} = \int {\cal D}\zeta \, {\cal D}z_1 \, {\cal D} z_2 \,
\exp \left (- {\tilde S}_{12} \right)
\label{tz12}
\ee
and involving only the quadratic in these variables part of the action
(\ref{s12})
\bea
&\dst \tilde {S} _{12} = \f {4 \pi} {3} \eps \, L^3  + {16 \, \pi \, \mu^3 \over
3 \, \eps^2} + & {\m \over 2} \, \int d \Omega \, g^{ab} \p_a \zeta \p_b \zeta +
{\e \over 2} \, \int_R^L r^2 \, dr \,d \Omega \, g^{ij} \, \p_i z \, \p_j z~.
\label{ts12}
\eea
In the same quadratic approximation the partition function ${\cal
Z}_1$ for the trivial configuration is given by
\be
{\cal Z}_1= \int {\cal D} z \, \exp \left (- S_1 \right )
\label{z1}
\ee
with $S_1$ given by Eq.(\ref{s1}) and the integral running over all the
functions vanishing at the space-time boundary: $z(L, \theta, \vp)=0$.

It was shown in \cite{mv08} that the calculation of the ratio of this type of
the partition functions can be reduced to a calculation of partition functions
associated with boundary only. It is clear what is meant by the boundary
partition function for bounce configuration. It is possible to define a similar
object for the flat wall configuration in the following way. Although there is
no relation between the flat wall configuration and the sphere with radius $ R =
\f {2 \, \m} {\e} $, one can calculate the partition function $ \mathcal {Z} _
1$ by first fixing the transverse variable $z$ at $r=R$: $z(R, \theta, \vp)=
\zeta(\theta,\vp)$ and separating the integration over the bulk variables, hence
introducing by hands the boundary for trivial configuration. As a result on gets
the ratio in the form
\be
\f {{\cal Z}_ {12} } {{\cal Z}_1} = \f {{\cal Z}_{12 (\rm boundary)}} {{\cal
Z}_{1 (\rm boundary)}},
\ee
with boundary partition functions given by
\be
\mathcal{Z}_ {12 (\rm boundary)} = \int {\cal D} \z \, \exp \l[ - {\m \over 2}
\, \int d \Omega \, \zeta \, \Delta ^ { (2) } \zeta + {\e R^2 \over 2} \, \int d
\Omega \, \zeta \, \p_r z_c \Big | _ { r = R } \r],
\label{z12b}
\ee
and
\be
\mathcal{Z}_ {1 (\rm boundary)} = \int {\cal D} \z \, \exp \l[{\e R^2 \over 2}
\, \int d \Omega \, \zeta \, \p_r z_c \Big | _ { r = R } \r],
\label{z1b}
\ee
with the function $z_c(r,\t,\vp)$ satisfying the Laplace equation $\Delta \,
z_c=0$ with the boundary conditions
\be
z _c (R, \t, \vp) = \z (\t, \vp), ~~~~ z _c (r = L)=0.
\label{zcbc}
\ee
The operator $\Delta ^ { (2) }$ is the angular part of the Laplace operator in
$3$d (the Laplace operator on a sphere).

One can find the complete set of these functions by expanding the boundary
function $\z (\t, \vp)$ in the series of spherical harmonics
\be
\z(\t, \vp) = \sum _ {l , m} A _ {l m} Y _ {l m} (\t, \vp) ~ .
\ee
In each of the partial waves  these functions are then found as
\be
z_c(r, \t, \vp) = A _ {l m} \f {R ^ {l+1}} {r ^ {l+1}} \, Y _ {l m} (\t, \vp) ~
.
\ee
Substituting these functions to the equations (\ref{z12b}) and (\ref{z1b}) and
performing integration over the amplitudes
$A _ {l m}$ yields
\be
\mathcal{Z}_ {12 (\rm boundary)} = {\cal N} \prod _ { l=0 } ^ { \infty }
\l( \f {1} {\m\, l\, (l+1) + \e \, R\, (l+1)} \r) ^ { (2 l + 1) /2 }
\ee
and
\be
\mathcal{Z}_ {1 (\rm boundary)} = {\cal N} \prod _ { l=0 } ^ { \infty }
\l( \f {1} {\e \, R\, (l+1)} \r) ^ { (2 l + 1) /2 }.
\ee

\section{Regularization}
Clearly, each of the formal expressions (\ref{z12b}) and (\ref{z1b})
contains a divergent product, and their ratio is also ill defined, so that our
calculation requires a regularization procedure that would cut off the
contribution of harmonics with large $l$. In order to perform such
regularization we use the standard Pauli-Villars method and introduce
regulator fields $Z_ \a$ with the weight factors $ C_\a $ such that
\be
\sum_\a C_\a = 1, ~~~~~ \sum _\a C_\a M _\a ^ n = 0,
\ee
for any $n$ less then some finite number. The action corresponding to the
quadratic part of the Nambu-Goto expression (\ref{NG}) for small $Z_ \a$:
\bea
&\dst \tilde {S} _{12} = {\m \over 2} \, \int d \Omega \, g^{ab} \p_a \zeta_ \a
\p_b \zeta_ \a  + {\e \over 2} \, \int_R^L r^2 \, dr \,d \Omega \, \l( g^{ij} \,
\p_i Z_ \a \, \p_j Z_ \a + M_ \a ^ 2 Z_ \a ^2 \r) ~.
\label{ts12r}
\eea
with $M _\a$ being each regulator mass, which physically should be understood as
satisfying the condition $M _\a \ll M_0$ and still being much larger than the
relevant scale in the discussed problem, in particular $M_ \a R \gg 1$.
The regularized expression for the ratio of the boundary terms in ${\cal
Z}_{12}$ and ${\cal Z}_{1}$ thus takes the form
\be
{{\cal Z}_{12 (\rm boundary)} \over {\cal Z}_{1 (\rm boundary)}} \longrightarrow
{\cal R}=
\left [ {{\cal Z}_{12 (\rm boundary)} \over {\cal Z}^{(R)}_{12 (\rm boundary)}}
\right ] \, \left [ {{\cal Z}_{1 (\rm boundary)} \over {\cal Z}^{(R)}_{1 (\rm
boundary)}} \right ]^{-1}~,
\label{reg}
\ee
where we introduced the notation ${\cal R}$ for the regularized ratio, and the
regulator partition functions ${\cal Z}^{(R)}_{12 (\rm boundary)}$
and ${\cal Z}^{(R)}_{1 (\rm boundary)}$ are determined by the same expressions
as in Eqs.(\ref{z12b}) and (\ref{z1b}) with function
$z_ {c}$ being replaced regulator functions counterparts $Z_ {\a c}$
which still satisfy the boundary conditions similar to
(\ref{zcbc}):
\be
Z_{\a c} (R, \theta, \vp)= \zeta_ \a (\theta, \vp)~,
\label{zrbc}
\ee
but are the solutions of the Helmholtz rather than the Laplace equation $(\Delta
-
M _\a ^2) Z _ \a=0$.
The solutions of the Helmholtz equation fall off exponentially at the scale
determined by $M _\a$, and for our purposes only the behavior near the sphere
$r=R$
is needed. For this reason we write the equation for the radial part of the
$l$-th angular harmonic as
\be
{Z}_{ l}''+{2 \over r} \, {Z}_{ l}' - {l (l+1)  \over r^2} \, {Z}_{l} - M^2 \,
{Z}_{ l} = 0~,
\label{zneq}
\ee
where for a time being we have omitted the indices $\a$ and $c$ for regulators.
Introducing new function $\tilde{Z}_{l}$:
$Z_l(r)=\sqrt{ {R} / {r} } \, \tilde{Z} _l(r)$ we can rewrite previous equation
in the form
\be
\tilde{Z}_{l }''+{1 \over r} \, \tilde{Z}_{l}' - {(l+1/2) ^2 \over r^2} \,
\tilde{Z}_{l } - M^2 \, \tilde{Z}_{l } = 0~.
\ee
One can now write the radial coordinate as $r=R+x$, and treat the parameter
$(x/R)$ as small, since the scale for the variation of the solution is $x \sim
1/\sqrt{M^2+(l+1/2)^2/R^2}$. This approach yields an expansion of the regulator
action
associated with the boundary at $r=R$ in powers of $1/\sqrt{(MR)^2 +
(l+1/2)^2}$. With
the accuracy required in the present calculation, the (normalized to one at
$r=R$) solution to Eq.(\ref{zneq}) is found in the first order of expansion in
$(x/R)$ as
\be
\tilde{Z}_l (R+x)=\sqrt { \f {R} {r} } \left ( 1- {1 \over 2} \, {(M R)^2 \over
(MR)^2+(l+1/2)^2} \, {x \over R}
\right ) \, \exp \left ( -\sqrt{(MR)^2+(l+1/2)^2}
 \, {|x| \over R} \right )~.
\label{znsol}
\ee
Using the form of the solutions for the harmonics of the regulator field given
by Eq.(\ref{znsol}) and the expressions (\ref{z12b}) and (\ref{z1b}), one
can
write the regularized ratio of the boundary partition functions (\ref{reg}) as
\bea
\mathcal {R} & = & \prod_{l=0}^\infty \l[ \f { l\, (l+1) + 2 \sqrt{M^2_\a \,R^2
+ \l(l+\f {1} {2} \r)^2} +
1} { (l+1) \, (l+2)} \r] ^ { {(2l+1) \, C_\a} / {2} } \times \nonumber \\
&& \prod_{l=0}^\infty \l[ \f {l + 1} { \sqrt{M^2_\a \,R^2 + \l(l+\f {1} {2}
\r)^2} +
\f {1} {2} } \r] ^ { {(2l+1) \, C_\a} / {2} } \times
\nonumber \\
&& \prod_{l=0}^\infty \l[ 1 + \f {M_ \a ^2 R^2} { \l( M^2_\a \,R^2 + \l(l+\f {1}
{2} \r)^2 \r) \,
\l( l\, (l+1) + 2 \sqrt{M^2_\a \,R^2 + \l(l+\f {1} {2} \r)^2} + 1 \r)} \r] ^ {
{(2l+1) \, C_\a} / {2} } ~,
\label{crprod}
\eea
where we took into account that $R = 2 \, \m / \e$.
\section{Calculating the products}
Each of the products in Eq.(\ref{crprod}) is finite at a finite $M$ and can be
calculated separately. Instead of calculating the product directly, we can use
the relation
\be
\prod_l F_l = \exp \l( \sum_l \ln F_l \r),
\ee
and calculate the sum. We start from the third product. The expression under the
sign of product is of the form
$1+g(l)$, with $g(l)$ close to $0$ for any $l$, since it behaves as $M^{-1}_ {
\a }$. Hence we can leave only the first term in the expansion of the logarithm
$\ln ( 1+x ) = x + O (x^2)$. Thus, we need to find the sum
\be
S_3 = \f {1} {2} \sum_ {l = 0} ^ \infty \sum_ {\a} C_ \a \, (2l+1) \, {M_ \a ^2
R^2} { \l[ M^2_\a \,R^2 + \l(l+\f {1} {2} \r)^2 \r] ^ {-1} \,
\l[ l\, (l+1) + 2 \sqrt{M^2_\a \,R^2 + \l(l+\f {1} {2} \r)^2} + 1 \r] ^ {-1}}~.
\ee
The sums associated with the three products are readily calculable with the help
of the Euler-Maclaurin summation formula and  the result for the regularized
ratio has the form
\be
\mathcal{R} = \exp \l[ \f {1} {2} M^2 R^2 \ln M R + M R \ln M R + \ln M R \r]
\label{reg_r_M}
\ee
where $M^n \ln M = \sum _\a C _\a M ^n _\a \ln M _\a$, for any $n$. The
expression for $\mathcal{R}$ contains an essential dependence on the regulator
mass parameter $M$. We will show, however, that all such dependence in the phase
transition rate can be absorbed in renormalization of the parameter $\mu$ in the
leading semiclassical term. Although, it may appear that there is a problem with
the term $M R \ln M$, which is not proportional to $R^2$ (the area of the
interface) and thus is not the coefficient in front of tension $\m$. But this
behavior is only due to the particular dependence of the radius $R$ on $\m$ and
$\e$.

\section{Renormalization of $\mu$}
The parameter $\mu$ is defined in the action (\ref{NG}) as the coefficient in
front of the area of the interface between the world space of the wall and
empty space.
Generally this parameter gets renormalized by the quantum
corrections, and in order to find such renormalization at the level of first
quantum corrections, one needs to perform the path integration using the
quadratic part of the action around a configuration, in which the area of the
boundary is an arbitrary parameter. For a practical calculation of this effect
we consider a Euclidean space configuration, with the wall lying flat in $(x,y)$
plane,
and the interface being at $y=0$. Thus, the area of the world surface of the
boundary
is $XT$, with $T$ and $X$ being the size of the world volume of the wall in the
$(t,x)$ plane.
The Gaussian path integral over the transverse coordinates $z(t,x,y)$ is then
to be calculated. We use the notation $\zeta(t,x)$ for the transverse shift of
the boundary and
expand it in the Fourier series
\be
\zeta(t,x) = \sum _ {\vec {k} } a( \vec{k} ) \, e ^ { i \vec {k} \vec {x} }.
\ee
A similar expansion applies to the regulator boundary function $\zeta_R(t,x)$.
The part of the effective action associated with the boundary is determined by
the functions $z_c(t,x,y)$ for the transverse shift of the string and the
corresponding regulator functions $Z_c(t,x,y)$ that satisfy the equations
\be
\Delta z_c=0 ~~~{\rm and} ~~~(\Delta - M _ \a ^ 2) \,Z_c =0~,
\label{laphel}
\ee
and the boundary conditions
\be
z_c(t,x,0) = \zeta(t,x)~,~~~~Z_c(t,x,0)=\zeta _ R (t,x)
\label{bcf}
\ee
as well as
\be
z_c(t,x,\pm \infty)=Z_c(t,x,\pm \infty)=0.
\label{bcfy}
\ee
One
can readily find these functions for each harmonic of the boundary values
$\zeta$ and $\zeta_R$
\be
z_c^{\vec{k}}(x,t,y) = e ^ { - { \sqrt{k^2} } \, |y|  } \, e ^ { i \vec {k} \vec {x}
} ~,
\label{zcf}
\ee
and
\be
Z_c^{\vec{k}}(x,t) = \exp \left [ -|y| \, \sqrt{ k^2 + M _ \a ^ 2} \, \right ] \, e
^ { i \vec {k} \vec {x} }~.
\label{zrcf}
\ee
In order to separate the boundary effect in the path integral around the
considered configuration from the bulk effects, we again divide it by the path
integral around the configuration where the whole world volume is occupied by
the
wall. Such division results, as previously, in the cancellation of the bulk
contributions, and
the remaining part of the effective action associated with the boundary is
written in terms of the regularized path integral over the boundary function
$\zeta$ as
\bea
&&\mu_R \, \mathcal{A} = \mu \, \mathcal{A} - \nonumber \\
&&\ln {\int {\cal D} \zeta  \, \exp \left \{- {1 \over 2} \, \int d\Omega \left
[ \mu \,
g ^ {ab} \, \p_a \z \, \p_b \z - \e\, \zeta \, \p _ y z_c(x,t,y)|_{y \to +0}
\right ] \right \} \over
\int {\cal D} \zeta_R  \, \exp \left \{ -{1 \over 2} \, \int d \Omega \left [
\mu \,
g ^ {ab} \, \p_a \z_R \, \p_b \z_R - \e\, \zeta \, \p _ y Z_c(x,t,y)|_{y \to +0}
\right ] \right \} } - \nonumber \\
&& \ln  {\int {\cal D} \zeta  \, \exp \left \{ - \f {1} {2} \e \, \zeta \, \p _
y z_c(x,t,y)|_{y \to +0} \right \} \over  \int {\cal D} \zeta_R  \, \exp \left
\{ - \f {1} {2} \e \, \zeta _ R \, \p _ y Z_c(x,t,y)|_{y \to +0} \right \} }~,
\label{muet}
\eea
where $\mu_R=\mu + \delta \mu$ is the renormalized mass parameter.  The
correction to $\mu$ can thus be written in the form
\bea
\delta \m & = & \f {1} {2} \int \f {d^2\, k} {(2 \pi)^2} \l[ \ln \l( k^2 + \f
{\e} {\m} \sqrt {M^2 + k^2} \r) -
\ln \l( k^2 + \f {\e} {\m} \sqrt {k^2} \r) \r] - \nonumber \\
&& \f {1} {4} \int \f {d^2\, k} {(2 \pi)^2} \l[ \ln \l( M^2 + k^2 \r) -
\ln k^2 \r] = \nonumber \\
&& \f {M^2} {8 \pi} \ln M R + \f {M \e} {8 \pi \m} \ln M R + \f {\e^2} {16
\pi\m^2} \ln M R.
\label{reg_m}
\eea
It should be mentioned here, that result for the renormalized parameter $\m_R$
would not change if we considered the wall not with a topology of a plane, but
rather of a cylinder, namely, considering periodic boundary conditions in one
direction.

\section{Results and conclusion}
Collecting all terms together we find the rate of the process. It is clear that
the result for each $d-3$ transverse dimensions factorizes, thus we have the
expression for the rate in the form
\be
\f {d \Gamma} {d A} = \f {\mathcal {C}} {\e ^ {7/3}} \, \mathcal{R} ^ {d-3} \,
\exp \l( - {16 \, \pi \, \mu^3 \over 3 \, \eps^2} \r) ,
\ee
where $\m$ is bare (non-renormalized) tension, and regularized ratio
$\mathcal{R}$ given by (\ref{reg_r_M}). Taking into account that each of the
transverse dimensions contributes additively to $\delta \mu$ and the interface
is a sphere with area
\be
\mathcal{A} = 4 \pi R ^ 2,
\ee
and expressing the bare $\mu$ through the renormalized one: $\mu=\m_R-\delta
\mu$, one readily finds that the dependence on the regulator mass M cancels in
the transition rate, and one arrives at the formula given by Eq.(\ref{rate}).

Thus we obtained the result similar to the decay of a string, when the effect of
all extra transverse dimensions results only in the renormalization of parameter
$\m$ associated with the interface. It should be mentioned, however, that the
result for a string was, actually, obtained for a transition between two states
of a string with different tensions. Here we considered decay of a wall
(transition into nothing). For the calculation used it is impossible to preserve
finite terms, but only proportional to some power of the regulator mass
parameter $M$.

Having calculated the probability rate for a decay of one- and two- dimensional
objects, e.g. string and domain wall, it is tempting to assume that the
situation is somewhat similar for the decay of an object of arbitrary
dimensionality. But it is not true already for the decay of tree- and four-
dimensional objects, where the dependence of the result on regulator mass is
substantial even after renormalization of a parameter associated with an
interface. That dependence demands introduction of new terms into initial
action, which corresponds to nonrenormalizibility of the effective `low-energy'
theory.

\section*{Acknowledgments}
This work is supported  in part by the DOE grant DE-FG02-94ER40823.

\end{document}